\documentclass[11pt,dvips]{article}
\usepackage{eurosym}
\usepackage{graphicx}
\usepackage{amsmath}
\usepackage{amssymb}
\usepackage{latexsym}
\usepackage{color}
\usepackage{epstopdf}

\begin{document}

\thispagestyle{empty}

\begin{center}
\vspace{1.8cm}

{\Large \textbf{Distributing long-distance trust in optomechanics}}

\vspace{0.7cm}

\textbf{Jamal El Qars}{\footnote{%
email: \textsf{j.elqars@uiz.ac.ma}}}

\vspace{0.3cm}

\textit{LMS3E, Faculty of Applied Sciences, Ibn Zohr University, Agadir,
Morocco}\\[0.3em]

\vspace{1.1cm} \textbf{Abstract}
\end{center}



Quantum steering displays an inherent asymmetric property that differs from
entanglement and Bell nonlocality. Besides being of fundamental interest,
steering is relevant to many asymmetric quantum information tasks. Here, we
propose a scheme to generate and manipulate Gaussian quantum steering
between two spatially distant mechanical modes of two optomechanical
cavities coupled via an optical fiber, and driven by blue detuned lasers. In
the unresolved sideband regime, we show, under realistic experimental
conditions, that strong asymmetric steering can be generated between the two
considered modes. Also, we show that one-way steering can be achieved and
practically manipulated through the lasers drive powers and the temperatures
of the cavities. Further, we reveal that the direction of one-way steering
depends on the sign of the difference between the energies of the mechanical
modes. Finally, we discuss how to access the generated steering. This work
opens up new perspectives for the distribution of long-distance trust which
is of great interest in secure quantum communication.


\section{Introduction}

In their seminal paper 1935 \cite{epr}, Einstein, Podolski, and Rosen (EPR) have considered two particles in a pure entangled state to illustrate why they questioned the incompatibility between the concept of local causality and the completeness of quantum mechanics. In fact, they pointed out that a measurement performed on one particle induces an apparent \textit{nonlocal} collapse of the quantum state of the other, which was unacceptable for them. The EPR paradox provoked an interesting analysis from Schr\"{o}dinger \cite{es1,es2}, who introduced the aspect of steering to describe the spooky action-at-a-distance discussed by EPR.

Within mixed states, quantum nonlocality can arise in various incarnations, i.e., Bell nonlocality \cite{bell}, quantum steering \cite{uola}, and entanglement \cite{horo}. In terms of violations of local hidden state model, EPR steering was defined by Wiseman and co-workers \cite{wiseman} as a form of quantum correlations that sets between Bell nonlocality \cite{bell} and entanglement \cite{horo}, and which allows one observer to remotely affect (steer) the state of another distant observer via local measurements \cite{dyang}. Importantly, Wiseman \textit{et al}. \cite{wiseman} showed the hierarchy of the three kind of quantum nonlocal states, that is, steerable states are a subset of the entangled states and a superset of Bell nonlocal states \cite{wiseman}. Put in other word, a bipartite state $\hat{\varrho}_{xy}$ that displays Bell nonlocality is steerable in both directions $x\rightleftarrows y$, whereas, EPR steering at least in one direction implies that the state is entangled \cite{uola}. The reverse scenario is not always true, meaning that not all entangled states are steerable, and not every steerable state exhibits Bell nonlocality \cite{costa}.

From a quantum information perspective \cite{uola}, quantum steering
corresponds to the task of verifiable entanglement distribution by an
untrusted party, i.e., if Alice and Bob share a quantum bipartite state
which is steerable, say, from Alice to Bob, therefore, Alice can convince
Bob, who does not trust her, that the shared state is entangled through
local measurements and classical communication.

On the basis of the inferred quadrature variances of light fields, Reid
proposed practical criteria for witnessing the existence of the EPR paradox
\cite{reid}, which were experimentally violated in \cite{ou1,ou2,ou4}.
Especially, it has been proven that violation of the Reid criteria, under
Gaussian measurements, demonstrates EPR steering \cite{wiseman}. Later,
various criteria which are explicitly concerned with EPR steering were
proposed \cite{senq}, where each criterion depends on the size of the
studied system and the detection method \cite{PRX}.

Besides, the quantification of EPR steering has attracted considerable
attention in the past decade, which results in miscellaneous useful
measures, e.g., steering weight \cite{sw} and steering robustness \cite{sr},
where both measures can be evaluated only by means of a semidefinite program
\cite{costa}. Within the Gaussian framework \cite{gs}, Kogias \textit{et al.}
\cite{kogias} introduced a computable measure of steering for arbitrary
bipartite Gaussian states. Essentially, they provided an operational
connection between the proposed measure and the key rate in one-sided
device-independent quantum key distribution (1SDI-QKD) \cite{qcry}.

Unlike entanglement and Bell nonlocality, EPR steering is intrinsically
asymmetric, i.e., a bipartite state $\hat{\varrho}_{xy}$ may be steerable
from $x \rightarrow y$, but not vice versa. Thus, one distinguishes, (%
\textit{i}) no-way steering where the state $\hat{\varrho}_{xy}$ is
nonsteerable neither from $x\rightarrow y$ nor in the reverse direction, (%
\textit{ii}) two-way steering where $\hat{\varrho}_{xy}$ is steerable in
both directions $x\rightleftarrows y$, and (\textit{iii}) one-way steering,
in which the steerability is authorized solely from $x\rightarrow y$ or $y
\rightarrow x$. It is now believed that the key ingredient of asymmetric quantum information tasks is EPR steering \cite{uola}, which has been
recognized as the essential resource for 1SDI-QKD \cite{qcry}, secure
quantum teleportation \cite{sqt1,sqt2,ram1}, subchannel discrimination \cite{sr},
quantum secret sharing \cite{qsc}, randomness generation \cite{rg}, and
secure quantum communication \cite{sqc,pseu}.

EPR steering has been investigated theoretically as well as experimentally
in various continuous-variable systems. Proposals include, cavity
magnomechanical systems \cite{mg}, Gaussian cluster states \cite{cluster},
optical systems \cite{photonic}, and nondegenerate three level laser system
\cite{elqars}. In particular, based on cavity optomechanics \cite{aspelmeyer}%
, a number of schemes for EPR steering generation were proposed in various
two-mode Gaussian states \cite{rui,guo,sun,zhang,zhu,tansun,dengli}. These,
however, cannot be used to implement long-distance secure quantum
communication, since the two considered modes belong to the same cavity.

The aim of this paper is to investigate, using experimentally feasible
parameters, the generation and manipulation of stationary Gaussian EPR
steering between two mechanical modes of two spatially separated Fabry-P\'{e}%
rot cavities coupled by an optical fiber. Our work goes beyond the proposals
put forward in Refs. \cite{jamal,jing}, where an additional squeezed light
source to drive the two cavities was considered.

Optomechanics involves hybrid coupling between optical and mechanical
degrees of freedom by means of radiation pressure \cite{aspelmeyer}. Over
the last few years, cavity optomechanics has emerged as a very promising
platform for demonstrating various quantum features. The important
achievements in this realm encompass preparing entangled states between
photons and phonons \cite{palomaki,vitali}, cooling the fundamental
vibrational mode to the ground state \cite{clerk}, creation of macroscopic
Schr\"{o}dinger's cat' states \cite{cat}, observing the radiation pressure
shot-noise \cite{teufl}, and squeezing effect \cite{painter}.

The remainder of this paper is organized as follows. In Sec. \ref{sec2}, we
introduce the basic optomechanical setup at hand. Next, using the standard Langevin formalism, we derive the covariance matrix describing the
whole system at the steady state. In Sec. \ref{sec3}, we quantify and study
Gaussian quantum steering between two spatially separated mechanical modes.
Also, we discuss how the generated mechanical steering can be experimentally
measured. In Sec. \ref{sec4}, we discuss a way to access experimentally the generated optomechanical steering. Finally, in Sec. \ref{sec5} we draw our conclusions.

\section{The system and its covariance matrix\label{sec2}}

\begin{figure}[tbh]
\centerline{\includegraphics[width=5.5cm]{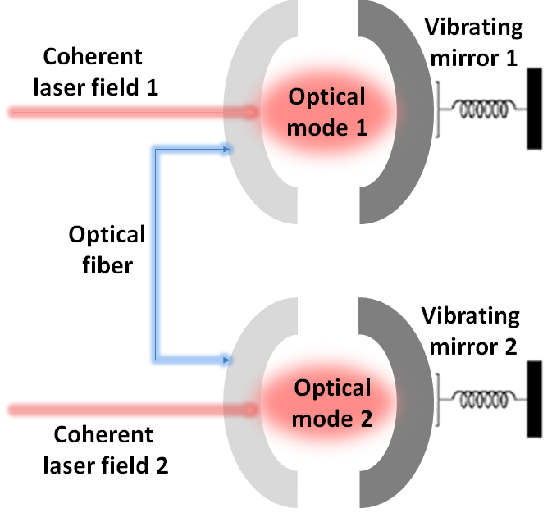}}
\caption{Schematic illustration of two spatially separated Fabry-P\'{e}rot
cavities coupled to each other by an optical fiber, and driven by two
coherent lasers fields. As a consequence of indirect coupling mediated by
the two optical modes, the two vibrating mirrors become quantumly
correlated. }
\label{f1}
\end{figure}

In Fig. \ref{f1}, we consider a double-cavity optomechanical system, where
each cavity comprises two mirrors. The first one is fixed and partially
transmitting, while the second is movable and perfectly reflective. The $j%
\mathrm{th}$ cavity, having equilibrium length $\ell _{j}$ and decay rate $%
\kappa _{j}$, is driven by a coherent laser field of power ${\wp }_{j}$ and
frequency $\omega _{L_{j}}$. In a frame rotating with the frequency $\omega
_{L_{j}}$, the system can be described by the Hamiltonian \cite{FNori}
\begin{equation}
\mathcal{\hat{H}}=\sum_{j=1}^{2}\Big(\hbar \left( \omega _{c_{j}}-\omega
_{L_{j}}\right) \hat{o}_{j}^{\dag }\hat{o}_{j}+\frac{\hbar \omega _{m_{j}}}{2%
}(\hat{q}_{j}+\hat{p}_{j})-\hbar g_{j}\hat{o}_{j}^{\dag }\hat{o}_{j}\hat{q}%
_{j}+\mathrm{i}\hbar \mathcal{E}_{j}(\hat{o}_{j}^{\dag }-\hat{o}_{j})+\hbar
\zeta \hat{o}_{j}^{\dag }\hat{o}_{3-j}\Big),  \label{e1}
\end{equation}%
where the first term describes the free Hamiltonian of the $j\mathrm{th}$
intracavity laser field modeled as a single optical mode with annihilation
operator $\hat{o}_{j}$ and frequency $\omega _{c_{j}}$. The second term
describes the free Hamiltonian of the $j\mathrm{th}$ movable mirror modeled
as a single mechanical mode with position-$\hat{q}_{j}$ and momentum-$\hat{p}%
_{j}$ operators, an effective mass $\mu _{j}$, frequency $\omega _{m_{j}}$,
and damping rate $\gamma _{j}$. The third term comes from the optomechanical
coupling, via the radiation pressure force, between the $j\mathrm{th}$
mechanical mode and the $j\mathrm{th}$ optical mode, with $g_{_{j}}=\frac{%
\omega _{c_{j}}}{\ell _{j}}\sqrt{\frac{\hbar }{\mu _{_{j}}\omega _{m_{j}}}}$%
. The fourth term describes the coupling between the $j\mathrm{th}$ input
laser and the $j\mathrm{th}$ optical mode, with $\mathcal{E}_{j}=\sqrt{\frac{%
2\kappa _{j}{\wp }_{j}}{\hbar \omega _{L_{j}}}}$. The last term corresponds
to the coupling between the two optical modes $\hat{o}_{1}$ and $\hat{o}_{2}$%
, with coupling strength $\zeta $.

Since the system at hand is affected by dissipation and noise, then its
dynamics can be described by the following quantum Langevin equation \cite%
{Clerk23}, i.e., $\frac{d\mathcal{\hat{O}}}{dt}=\frac{1}{\mathrm{i}\hbar }%
\left[ \mathcal{\hat{O}},\mathcal{\hat{H}}\right] +$$\mathcal{\hat{D}}+%
\mathcal{\hat{N}}$ (for $\mathcal{\hat{O}\equiv }\hat{o}_{j},\hat{q}_{j}$, $%
\hat{p}_{j}$), where $\mathcal{\hat{D}}$ and $\mathcal{\hat{N}}$
characterize the dissipation and noise, respectively. For the Hamiltonian (%
\ref{e1}), we obtain
\begin{eqnarray}
\frac{d\hat{o}_{j}}{dt} &=&-\left( \kappa _{j}+i\Delta _{0_{j}}\right) \hat{o%
}_{j}+\mathrm{i}g_{j}\hat{o}_{j}\hat{q}_{j}-\mathrm{i}\zeta \hat{o}_{3-j}+%
\mathcal{E}_{j}+\sqrt{2\kappa _{j}}\hat{o}_{j}^{in},  \label{e2} \\
\frac{d\hat{q}_{j}}{dt} &=&\omega _{m_{j}}\hat{p}_{j},  \label{e3} \\
\frac{d\hat{p}_{j}}{dt} &=&-\omega _{m_{j}}\hat{q}_{j}+g_{j}\hat{o}%
_{j}^{\dag }\hat{o}_{j}-\gamma _{j}\hat{p}_{j}+\hat{\xi}_{j},  \label{e4}
\end{eqnarray}%
where $\Delta _{0_{j}}=\omega _{c_{j}}-\omega _{L_{j}}$ is the $j\mathrm{th}$
laser detuning \cite{Clerk23}. In Eq. (\ref{e2}), $\hat{o}_{j}^{in}$ denotes
the $j\mathrm{th}$ zero mean vacuum radiation input noise with the
correlation function $\langle \hat{o}_{j}^{in}(t)\hat{o}_{j}^{in\dag
}(t^{\prime })\rangle =\delta (t-t^{\prime })$ \cite{Milburn}. While, in Eq.
(\ref{e4}) $\hat{\xi}_{j}$ is the zero-mean Brownian noise operator
affecting the $j\mathrm{th}$ mechanical mode. It turns out that quantum
effects can be reached only by using mechanical oscillators with high
mechanical quality factor $Q_{j}=\omega _{m_{j}}/\gamma _{j}\gg 1$ \cite%
{aspelmeyer}. In this limit, we have the following nonzero correlation
function \cite{Benguria}
\begin{equation}
\langle \hat{\xi}_{j}(t)\hat{\xi}_{j}(t^{\prime })+\hat{\xi}_{j}(t^{\prime })%
\hat{\xi}_{j}(t)\rangle /2=\gamma _{j}(2n_{\mathrm{th},j}+1)\delta
(t-t^{\prime }),
\end{equation}%
where $n_{\mathrm{th},j}=(e^{\hbar \omega _{m_{j}}/k_{B}T_{j}}-1)^{-1}$ is
the $j\mathrm{th}$ mean thermal phonon number. $T_{j}$ and $k_{B}$ are the $j%
\mathrm{th}$ mirror temperature and the Boltzmann constant, respectively.

Due to the nonlinearity of Eq. (\ref{e1}), the coupled quantum Langevin
equations [(\ref{e2})-(\ref{e4})] are not in general solvable analytically.
However, using bright inputs lasers, the linearization of these equations
around the steady-state values is justified, and the dynamics can be solved
exactly. Hence, by adopting the standard linearization method \cite{Milburn}%
, we decompose each operator ($\mathcal{\hat{O}\equiv }\hat{o}_{j},\hat{q}%
_{j}$, $\hat{p}_{j}$) as a sum of the steady-state value $\langle \mathcal{%
\hat{O}\rangle }$ and a fluctuation operator $\delta \mathcal{\hat{O}%
}$ with zero mean value, i.e., $\mathcal{\hat{O}=}$ $\langle \mathcal{\hat{O}%
\rangle +}\delta \mathcal{\hat{O}}$. By setting $\frac{d}{dt}=0$ in Eqs. [(%
\ref{e2})-(\ref{e4})] and solving the obtained equations, we get $\langle
\hat{q}_{j}\rangle =\frac{g_{j}\left\vert \langle \hat{o}_{j}\rangle
\right\vert ^{2}}{\omega _{m_{j}}}$, $\langle \hat{p}_{j}\rangle =0$, and $%
\langle \hat{o}_{j}\rangle =\frac{\mathcal{E}_{j}\left[ \kappa _{j}+\mathrm{i%
}\left( -\zeta +\Delta _{3-j}\right) \right] }{\zeta ^{2}+\kappa
_{j}^{2}-\Delta _{1}\Delta _{2}+\mathrm{i}\kappa _{j}(\Delta _{1}+\Delta
_{2})}$, with $\Delta _{j}$ $=$ $\Delta _{0_{j}}$ $-g_{j}\langle \hat{q}%
_{j}\rangle $ being the $j\mathrm{th}$ effective cavity detuning \cite%
{Clerk23}. Now, by inserting $\mathcal{\hat{O}=}$ $\langle \mathcal{\hat{O}%
\rangle +}\delta \mathcal{\hat{O}}$ into Eqs. [(\ref{e2})-(\ref{e4})], and
assuming that the two cavities are intensely driven, i.e., $\left\vert
\langle \hat{o}_{j}\rangle \right\vert \gg 1$, which allows us to safely
neglect the quadratic terms $\delta \hat{o}_{j}^{\dag }\delta \hat{o}_{j}$
and $\delta \hat{o}_{j}\delta \hat{q}_{j}$, we obtain
\begin{eqnarray}
\frac{d\delta \hat{o}_{j}}{dt} &=&-\left( \kappa _{j}+\mathrm{i}\Delta
_{j}\right) \delta \hat{o}_{j}+\mathrm{i}g_{j}\langle \hat{o}_{j}\rangle
\delta \hat{q}_{j}-\mathrm{i}\zeta \delta \hat{o}_{3-j}+\sqrt{2\kappa _{j}}%
\hat{o}_{j}^{in},  \label{e7} \\
\frac{d\delta \hat{q}_{j}}{dt} &=&\omega _{m_{j}}\delta \hat{p}_{j},
\label{e8} \\
\frac{d\delta \hat{p}_{j}}{dt} &=&-\omega _{m_{j}}\delta \hat{q}%
_{j}+g_{j}\langle \hat{o}_{j}\rangle \left( \delta \hat{o}_{j}^{\dag
}+\delta \hat{o}_{j}\right) -\gamma _{j}\delta \hat{p}_{j}+\hat{\xi}_{j}.
\label{e9}
\end{eqnarray}

We emphasize that Eq. (\ref{e9}) is obtained under the assumption that the
phase reference of the $j\mathrm{th}$ laser field is chosen such that $%
\langle \hat{o}_{j}\rangle $ is real. Next, introducing the $j\mathrm{th}$
optical mode quadratures $\delta \hat{x}_{j}=(\delta \hat{o}_{j}^{\dag
}+\delta \hat{o}_{j})/\sqrt{2}$ and $\delta \hat{y}_{j}=i(\delta \hat{o}%
_{j}^{\dag }-\delta \hat{o}_{j})/\sqrt{2}$ as well as the $j\mathrm{th}$
input noise quadratures $\delta \hat{x}_{j}^{in}=(\delta \hat{o}_{j}^{in\dag
}+\delta \hat{o}_{j}^{in})/\sqrt{2}$ and $\delta \hat{y}_{j}^{in}=i(\delta
\hat{o}_{j}^{in\dag }-\delta \hat{o}_{j}^{in})/\sqrt{2}$, and using Eqs. [(%
\ref{e7})-(\ref{e9})], one gets%
\begin{equation}
\frac{d\hat{u}(t)}{dt}=\mathcal{K}\hat{u}(t)+\hat{n}(t),  \label{e10}
\end{equation}%
where the transposes of the fluctuations vector $\hat{u}(t)$ and the noises
vector $\hat{n}(t)$ are, respectively,
\begin{eqnarray}
\hat{u}(t)^{\mathrm{T}} &=&\left( \delta \hat{q}_{1}(t),\delta \hat{p}%
_{1}(t),\delta \hat{q}_{2}(t),\delta \hat{p}_{2}(t),\delta \hat{x}%
_{1}(t),\delta \hat{y}_{1}(t),\delta \hat{x}_{2}(t),\delta \hat{y}%
_{2}(t)\right) ,  \label{e11} \\
\hat{n}(t)^{\mathrm{T}} &=&(0,\hat{\xi}_{1}(t),0,\hat{\xi}_{2}(t),\delta
\hat{x}_{1}^{in}(t),\delta \hat{y}_{1}^{in}(t),\delta \hat{x}%
_{2}^{in}(t),\delta \hat{y}_{2}^{in}(t)),  \label{e12}
\end{eqnarray}%
and the kernel $\mathcal{K}$ is given by
\begin{equation}
\mathcal{K=}\left(
\begin{array}{cccccccc}
0 & \omega _{m_{1}} & 0 & 0 & 0 & 0 & 0 & 0 \\
-\omega _{m_{1}} & -\gamma _{1} & 0 & 0 & G_{1} & 0 & 0 & 0 \\
0 & 0 & 0 & \omega _{m_{2}} & 0 & 0 & 0 & 0 \\
0 & 0 & -\omega _{m_{2}} & -\gamma _{2} & 0 & 0 & G_{2} & 0 \\
0 & 0 & 0 & 0 & \kappa _{1} & \Delta _{1} & 0 & \zeta \\
G_{1} & 0 & 0 & 0 & -\Delta _{1} & \kappa _{1} & -\zeta & 0 \\
0 & 0 & 0 & 0 & 0 & \zeta & \kappa _{2} & \Delta _{2} \\
0 & 0 & G_{2} & 0 & -\zeta & 0 & -\Delta _{2} & \kappa _{2}%
\end{array}%
\right) ,  \label{e13}
\end{equation}%
with
\begin{equation}
G_{j}=\sqrt{2}g_{j}\langle \hat{o}_{j}\rangle =\frac{2\omega _{c_{j}}}{\ell
_{j}}\sqrt{\frac{\kappa _{j}{\wp }_{j}[\kappa _{j}^{2}+\left( \Delta
_{3-j}-\zeta \right) ^{2}]}{\mu _{_{j}}\omega _{m_{j}}\omega _{L_{j}}[(\zeta
^{2}+\kappa _{j}^{2}-\Delta _{1}\Delta _{2})^{2}+\kappa _{j}^{2}(\Delta
_{1}+\Delta _{2})^{2}]}},  \label{gg}
\end{equation}%
being the $j\mathrm{th}$ effective coupling \cite{aspelmeyer}. The solution
of Eq. (\ref{e10}) writes \cite{vitali}
\begin{equation}
\hat{u}(t)=f(t)\hat{u}(0)+\int_{0}^{t}dsf(s)\hat{n}(t-s),  \label{e15}
\end{equation}%
with $f(t)=e^{\mathcal{K}t}$. The system under consideration is stable and
reaches its steady state when all the eigenvalues of the matrix $\mathcal{K}$
have negative real parts, so that $f(\infty )=0$. The stability conditions
can be deduced by applying the Routh-Hurwitz criterion \cite{RH}. But in our
four-mode Gaussian state case, they are quite involved, and cannot be
reported here.

The operators $\hat{o}_{j}^{in}$ and $\hat{\xi}_{j}$ are zero-mean quantum
Gaussian noises and the dynamics is linearized. Then, the quantum steady
state of the fluctuations is a zero-mean Gaussian state, entirely
characterized by its $8\times8$ covariance matrix $\mathcal{V}_{ii^{\prime
}}=(\langle \hat{u}_{i}(\infty )\hat{u}_{i^{\prime }}(\infty )+\hat{u}%
_{i^{\prime }}(\infty )\hat{u}_{i}(\infty )\rangle )/2$, where $\hat{u}%
(\infty )$ is the fluctuations vector at the steady state \cite{vitali}.
When the system is stable and using Eq. (\ref{e15}), we obtain
\begin{equation}
\mathcal{V}_{ii^{\prime }}=\sum_{k,k^{\prime }}\int_{0}^{\infty
}ds\int_{0}^{\infty }ds^{\prime }f_{^{ik}}(s)f_{^{i^{\prime }k^{\prime
}}}(s^{\prime })\Phi _{^{kk^{\prime }}}(s-s^{\prime }),  \label{e17}
\end{equation}%
where $\Phi _{^{kk^{\prime }}}(s-s^{\prime })=(\langle \hat{n}_{k}(s)\hat{n}%
_{k^{\prime }}(s^{\prime })+\hat{n}_{k^{\prime }}(s^{\prime })\hat{n}%
_{k}(s)\rangle )/2=D_{^{kk^{\prime }}}\delta (s-s^{\prime })$ denotes the
matrix of stationary noise correlation functions \cite{vitali}. Using the
correlation properties of the operators $\hat{o}_{j}^{in}$ and $\hat{\xi}%
_{j} $, one can show that $D=D_{\gamma }\oplus D_{\kappa }$, with $D_{\gamma
}=\mathrm{diag}\left[ 0,\gamma_{1} \left( 2n_{\mathrm{th,1}}+1\right)
,0,\gamma_{2} \left( 2n_{\mathrm{th,2}}+1\right) \right] $ and $D_{\kappa }=%
\mathrm{diag}\left[\kappa_{1},\kappa_{1},\kappa_{2},\kappa_{2}\right]$.
Hence, Eq. (\ref{e17}) becomes $\mathcal{V}=\int_{0}^{\infty }dsf(s)Df(s)^{%
\mathrm{T}}$ which is equivalent to the Lyapunov equation \cite{jamal}
\begin{equation}
\mathcal{KV}+\mathcal{VK}^{\mathrm{T}}=-D.  \label{e18}
\end{equation}

The mechanical covariance matrix $\mathcal{V}_{m}$ of the two mechanical
modes, labelled as $m_{1}$ and $m_{2}$, can be deduced by tracing over the
optical elements in the general expression of $\mathcal{V}$. Then, we get
\begin{equation}
\mathcal{V}_{m}=\left(
\begin{array}{cc}
\mathcal{V}_{m_{1}} & \mathcal{V}_{m_{1}/m_{2}} \\
\mathcal{V}_{m_{1}/m_{2}}^{\mathrm{T}} & \mathcal{V}_{m_{2}}%
\end{array}%
\right),  \label{e19}
\end{equation}%
where the $2\times 2$ matrices $\mathcal{V}_{m_{1}}$ and $\mathcal{V}%
_{m_{2}} $ represent, respectively, the first and second mechanical modes,
while $\mathcal{V}_{m_{1}/m_{2}}$ describes the correlations between them.

\section{Gaussian quantum steering\label{sec3}}

It has been shown in \cite{wiseman} that a bipartite Gaussian state $\hat{%
\varrho}_{m_{1}m_{2}}$ with covariance matrix $\mathcal{V}_{m}$ is steerable
under Gaussian measurements performed on party $m_{1}$ if, and only if, the
inequality $\mathcal{V}_{m}+\mathrm{i}(\Omega _{m_{1}}\oplus \Omega
_{m_{2}})\geqslant 0$ is violated, where $\Omega _{m_{1}}=\left(
\begin{array}{cc}
0 & 0 \\
0 & 0%
\end{array}%
\right) $ and $\Omega _{m_{2}}$ $=\left(
\begin{array}{cc}
0 & 1 \\
-1 & 0%
\end{array}%
\right) $. This constraint yielded Kogias \textit{et al}. \cite{kogias} to
introduce a computable measure for quantifying the amount by which an
arbitrary bipartite Gaussian state $\hat{\varrho}_{m_{1}m_{2}}$ is steerable
under Gaussian measurements implemented on party $m_{1}$, i.e.,
\begin{equation}
\mathcal{G}^{m_{1}\rightarrow m_{2}}:=\max \left[ 0,-\sum\limits_{j:\bar{\nu}%
_{j}^{m_{2}}<1}\ln \left\{ {\bar{\nu}_{j}^{m_{2}}}\right\} \right],
\label{e20}
\end{equation}%
where $\{\bar{\nu}_{j}^{m_{2}}\}$ denotes the symplectic spectra of the
Schur complement $\mathcal{V}_{m_{2}}-\mathcal{V}_{m_{1}/m_{2}}^{\mathrm{T}}%
\mathcal{V}_{m_{1}}^{-1}\mathcal{V}_{m_{1}/m_{2}}$ of $m_{1}$ in the
covariance matrix $\mathcal{V}_{m}$ (\ref{e19}). The steering $\mathcal{G}%
^{m_{1}\rightarrow m_{2}}$ is monotone under Gaussian local operations and
classical communication, and it vanishes if the state described by $\mathcal{%
V}_{m}$ is nonsteerable under Gaussian measurements performed on party $%
m_{1} $. When $m_{1}$ and $m_{2}$ are two single modes, which we consider in
this paper, Eq. (\ref{e20}) acquires the simple form $\mathcal{G}%
^{m_{1}\rightarrow m_{2}}=\max \left[ 0,\frac{1}{2}\ln \frac{\det \mathcal{V}%
_{m_{1}}}{4\det \mathcal{V}_{m}}\right]$ \cite{kogias}. Similarly, quantum
steering in the reverse direction can be obtained by changing the roles of $%
m_{1}$ and $m_{2}$ in the expression of $\mathcal{G}^{m_{1}\rightarrow
m_{2}} $, i.e., $\mathcal{G}^{m_{2}\rightarrow m_{1}}=\max \left[ 0,\frac{1}{%
2}\ln \frac{\det \mathcal{V}_{m_{2}}}{4\det \mathcal{V}_{m}}\right] $. A
nonzero value of $\mathcal{G}^{m_{1}(m_{2})\rightarrow m_{2}(m_{1})}$ means
that the state $\hat{\varrho}_{m_{1}m_{2}}$ is steerable from $%
m_{1}(m_{2})\rightarrow m_{2}(m_{1})$ under Gaussian measurements performed
on mode $m_{1}(m_{2})$.

\begin{figure}[t]
\centerline{\includegraphics[width=0.4\columnwidth,height=4cm]{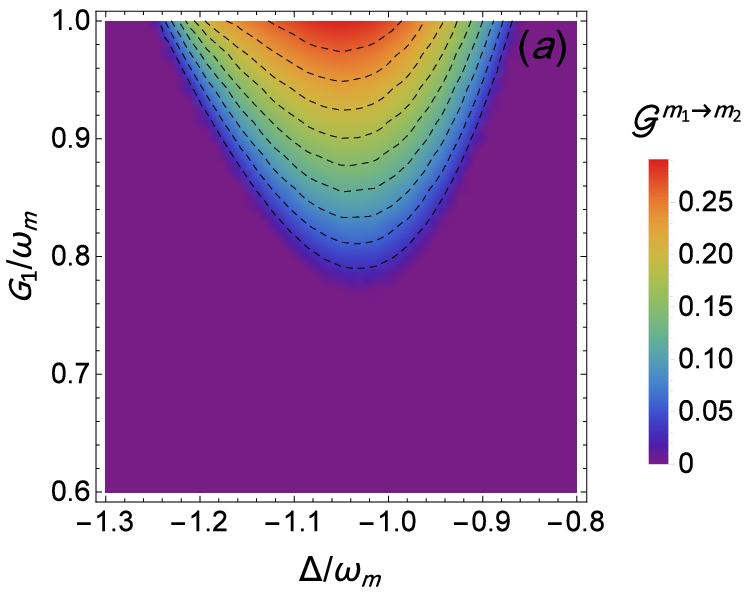}
\includegraphics[width=0.4\columnwidth,height=4cm]{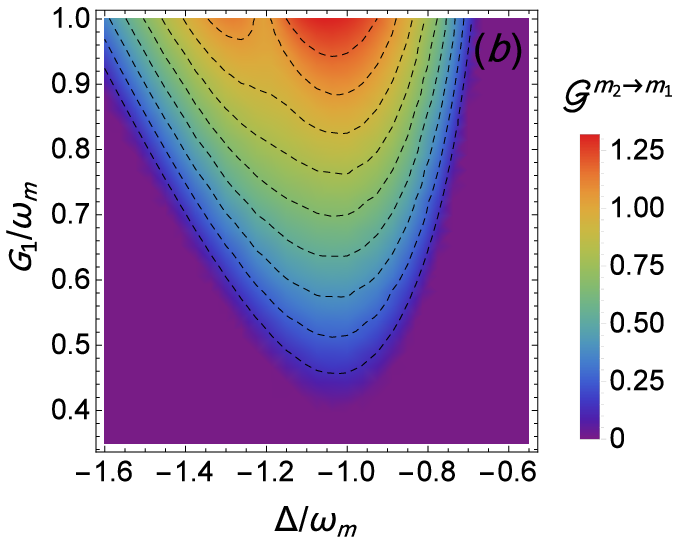}}
\centerline{\includegraphics[width=0.4\columnwidth,height=4cm]{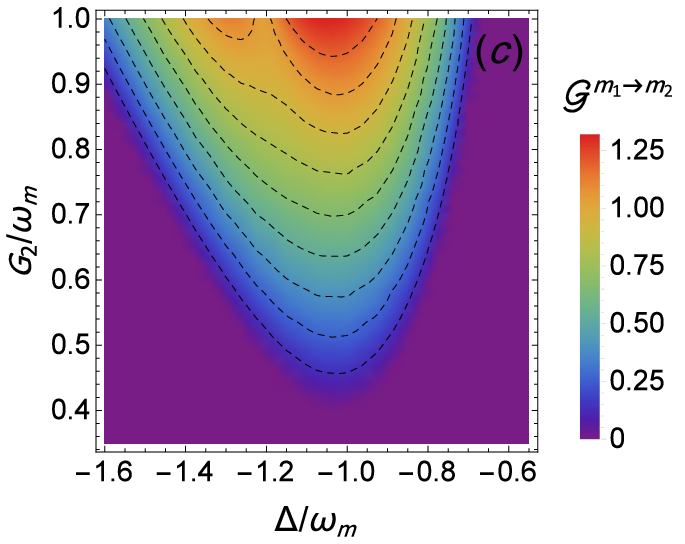}
\includegraphics[width=0.4\columnwidth,height=4cm]{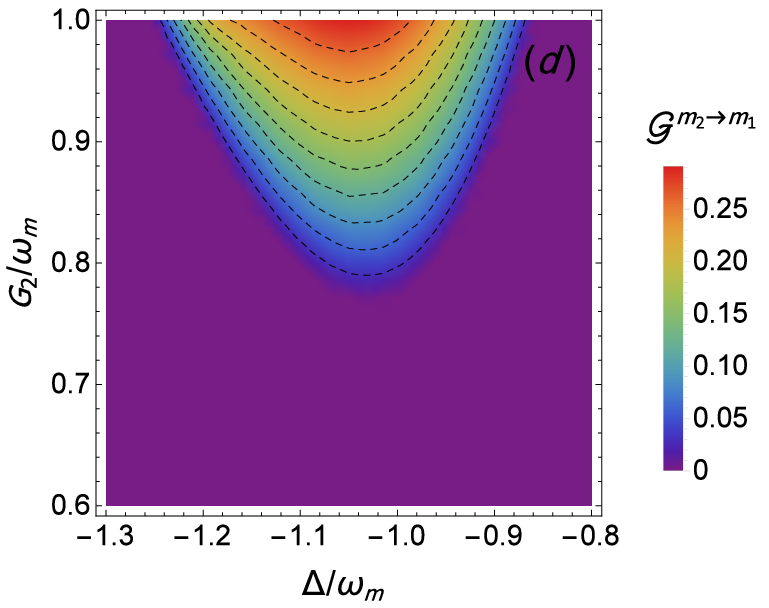}}
\caption{(a) and (b) show, respectively, the steerabilities $\mathcal{G}
^{m_{1}\rightarrow m_{2}}$ and $\mathcal{G}^{m_{2}\rightarrow m_{1}}$ versus
the normalized detuning $\Delta / \protect\omega _{m}$ and the normalized
optomechanical coupling $G_{1}/\protect\omega _{m}$ for $G_{2}=0.5G_{1}$.
(c) and (d) show, respectively, $\mathcal{G}^{m_{1}\rightarrow m_{2}}$ and $%
\mathcal{G} ^{m_{2}\rightarrow m_{1}}$ versus $\Delta /\protect\omega _{m}$
and $G_{2}/\protect\omega _{m}$ for $G_{1}=0.5G_{2}$. The other parameters
are $\protect\omega _{m}/2\protect\pi =947~\mathrm{kHz}$, $\protect\kappa /2%
\protect\pi =215~\mathrm{kHz}$, $\protect\gamma /2\protect\pi =140~\mathrm{Hz%
}$, $\protect\zeta /\protect\omega _{m}=5\times 10^{-2}$, and $n_{\text{th}%
,1}=n_{\text{th},2}=25$.}
\label{f2}
\end{figure}

For achieving asymmetric steering between the two mechanical modes $m_{1}$
and $m_{2}$, we introduce asymmetry between them by imposing $G_{1}\neq
G_{2} $. This can be practically realised by choosing identical parameters
for the two cavities except the drive lasers powers, i.e., $\wp _{1}\neq \wp
_{2}$. Moreover, for realistic estimation of the steerabilities $\mathcal{G}%
^{m_{1}\rightarrow m_{2}}$ and $\mathcal{G}^{m_{2}\rightarrow m_{1}}$, we
use parameters from the experiment \cite{Groblacher}. The movable mirrors,
having a mass $\mu =145~\mathrm{ng}$ and frequency $\omega _{m}=2\pi \times
947~\mathrm{kHz}$, are damped at rate $\gamma =2\pi \times 140~\mathrm{Hz}$.
The two cavities have equilibrium length $\ell =25~\mathrm{mm}$, decay rate $%
\kappa =2\pi \times 215~\mathrm{kHz}$, frequency $\omega _{c}=2\pi \times
5.26\times 10^{14}~\mathrm{Hz}$, and pumped by lasers of frequency $\omega
_{L}=2\pi \times 2.82\times 10^{14}~\mathrm{Hz}$.

Figs. \ref{f2}(a) and \ref{f2}(b) show, respectively, the steerabilities $%
\mathcal{G}^{m_{1}\rightarrow m_{2}}$ and $\mathcal{G}^{m_{2}\rightarrow
m_{1}}$ versus the normalized detuning $\Delta /\omega _{m}$ and the
normalized optomechanical coupling strength $G_{1}/\omega _{m}$ using $%
G_{2}=0.5G_{1}$. While, Figs. \ref{f2}(c) and \ref{f2}(d) show,
respectively, $\mathcal{G}^{m_{1}\rightarrow m_{2}}$ and $\mathcal{G}%
^{m_{2}\rightarrow m_{1}}$ versus $\Delta /\omega _{m}$ and $G_{2}/\omega
_{m}$ using $G_{1}=0.5G_{2}$. In the four cases, we used $n_{\text{th},1}=n_{%
\text{th},2}=25$ and $\zeta /\omega _{m}=5\times 10^{-2}$. As shown, without
external lasers, i.e., $\wp_{j}=0$, which, following Eq. (\ref{gg}), is
equivalent to $G_{j}=0$, no steering can be created between the modes $m_{1}$
and $m_{2}$ in both directions, namely $\mathcal{G}^{m_{1}\rightarrow m_{2}}=%
\mathcal{G}^{m_{2}\rightarrow m_{1}}=0$. Besides, by increasing gradually
the drive lasers powers $\wp _{1}$ and $\wp _{2}$, which in turn enhances
the optomechanical coupling strengths $G_{1}$ and $G_{2}$, the
steerabilities $\mathcal{G}^{m_{1}\rightarrow m_{2}}$ and $\mathcal{G}%
^{m_{2}\rightarrow m_{1}}$ appear and grow monotonously, reaching their
maximum around $\Delta /\omega _{m}\approx -1$ for $G_{1}/\omega _{m}\approx
1$ in [Figs. \ref{f2}(a) and \ref{f2}(b)], and for $G_{2}/\omega _{m}\approx
1$ in [Figs. \ref{f2}(c) and \ref{f2}(d)].

\begin{figure}[t]
\centerline{\includegraphics[width=0.4\columnwidth,height=4cm]{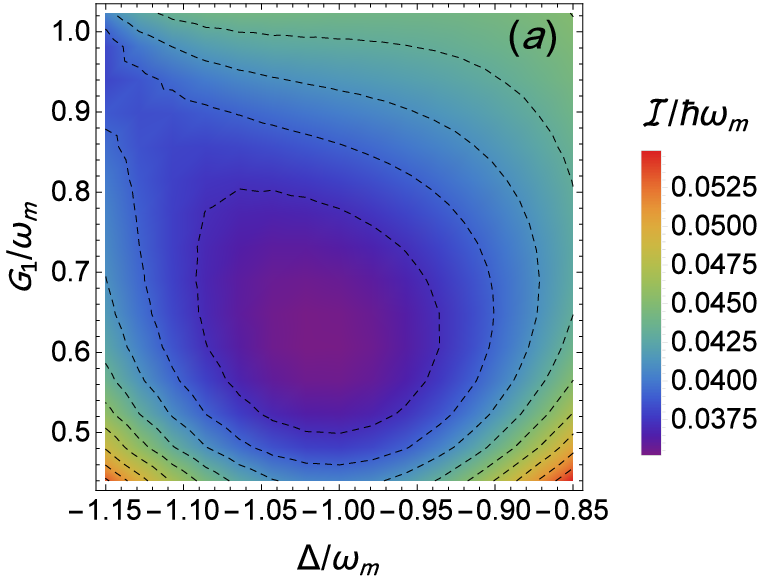}%
\includegraphics[width=0.4\columnwidth,height=4cm]{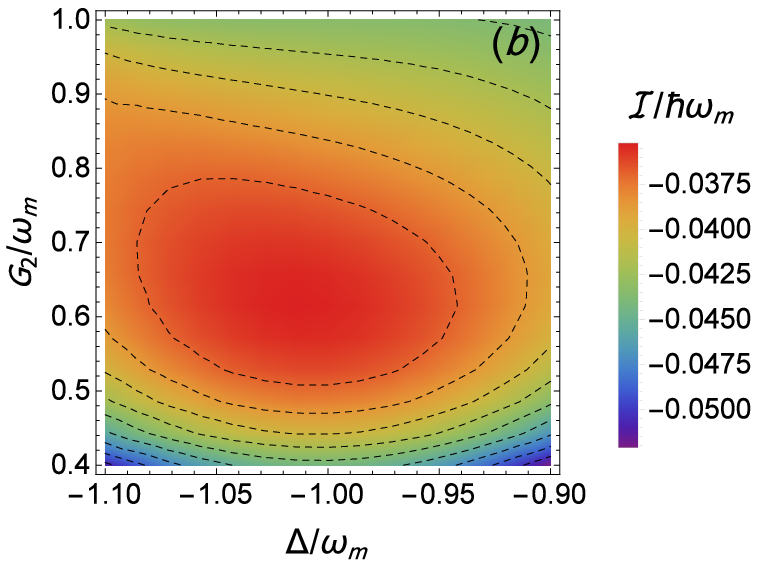}}
\caption{Plot of $\mathcal{I}$ (the difference between the energies of the
mechanical modes $m_{1}$ and $m_{2}$), in units of $\hbar\protect\omega_{m}$%
, against $\Delta/\protect\omega_{m}$ and $G_{1}/\protect\omega_{m}$ in (a),
and against $\Delta/\protect\omega_{m}$ and $G_{2}/\protect\omega_{m}$ in
(b). The parameters in (a) and (b) are, respectively, the same as in [Figs.
\protect\ref{f2}(a) and \protect\ref{f2}(b)] and [Figs. \protect\ref{f2}(c)
and \protect\ref{f2}(d)].}
\label{f3}
\end{figure}

Quite remarkably, Fig. \ref{f2} shows that the state $\hat{\varrho}%
_{m_{1}m_{2}}$ of the two modes $m_{1}$ and $m_{2}$ can display two-way
steering and even one-way steering. For example, [Figs. \ref{f2}(a) and \ref%
{f2}(b)] show that $\mathcal{G}^{m_{1}\rightarrow m_{2}}$ and $\mathcal{G}%
^{m_{2}\rightarrow m_{1}}$ are non-zero for $\Delta /\omega _{m}\approx -1$
and $G_{1}/\omega _{m}\geqslant 0.8$. This means that two-way steering can
be detected over a wide range of operating parameters, which is proven in
\cite{sqt2} to be a necessary resource needed for teleporting a coherent
state with fidelity beyond the no-cloning threshold. Moreover, for $\Delta
/\omega _{m}\approx -1$ and $0.45\leqslant G_{1}/\omega _{m}\leqslant 0.8$,
we have $\mathcal{G}^{m_{1}\rightarrow m_{2}}=0$ and $\mathcal{G}%
^{m_{2}\rightarrow m_{1}}>0$, meaning that the state $\hat{\varrho}%
_{m_{1}m_{2}}$ is one-way steerable from $m_{2}\rightarrow m_{1}$. The
different degree of steering observed between the directions $m_{1}
\rightarrow m_{2}$ and $m_{2} \rightarrow m_{1}$ is also proven to provide
the asymmetric guaranteed key rate achievable within a practical 1SDI-QKD
\cite{kogias}.

Importantly, [Figs. \ref{f2}(a) and \ref{f2}(b)] show that the state $\hat{%
\varrho}_{m_{1}m_{2}}$ is one-way steerable from $m_{2}\rightarrow m_{1} $
for $G_{1}/G_{2}>1$, while it is one-way steerable in the reverse direction $%
m_{1}\rightarrow m_{2}$ for $G_{1}/G_{2}<1$ as depicted in [Figs. \ref{f2}%
(c) and \ref{f2}(d)]. Then, we conclude that the direction of one-way
steering between the modes $m_{1}$ and $m_{2}$ could be merely manipulated
through the ratio  $G_{1}/G_{2}$ of the optomechanical coupling strengths,
and then through the ratio $\wp_{1}/\wp_{2}$ of the input lasers powers
following Eq. (\ref{gg}). This therefore provides an experimental flexible
and feasible way to control the direction of one-way steering, in comparison
with the loss-manipulating-method \cite{losses} that cannot be conveniently
adjusted within the experimental operations.

To better explain the behavior of the direction of one-way steering observed
in [Figs. \ref{f2}(a) and \ref{f2}(b)] and [Figs. \ref{f2}(c) and \ref{f2}%
(d)], we investigate the sign of the difference $\mathcal{I}=\mathcal{U}_{1}-%
\mathcal{U}_{2}$, where $\mathcal{U}_{j}=\frac{\hbar \omega _{m}}{2}(
\langle \delta \hat{q}_{j}^{2}\rangle+\langle \delta \hat{p}_{j}^{2}\rangle)$
is the mean energy of the $j$\textrm{th} mechanical mode. From now on, we
study $\mathcal{I}$ in units of $\hbar\omega_{m}$. Fig. \ref{f3}(a) shows
the difference $\mathcal{I}$ under the same circumstances of [Figs. \ref{f2}%
(a) and \ref{f2}(b)]. While, Fig. \ref{f3}(b) shows $\mathcal{I}$ under the
same circumstances of [Figs. \ref{f2}(c) and \ref{f2}(d)]. Manifestly, the
direction of one-way steering is strongly influenced by the sign of $%
\mathcal{I}$, i.e., in Fig. \ref{f3}(a) where $\mathcal{I}$ remains
positive, one-way steering is occurred from $m_{2}\rightarrow m_{1}$ in
[Figs. \ref{f2}(a) and \ref{f2}(b)]. Whereas, in Fig. \ref{f3}(b) where $%
\mathcal{I}$ remains negative, one-way steering is occurred in the reverse
direction $m_{1}\rightarrow m_{2}$ as illustrated in [Figs. \ref{f2}(c) and %
\ref{f2}(d)].

From a practical point of view, one-way steering from $m_{1}\rightarrow
m_{2} $ can be interpreted as follows: Alice (owning mode $m_{1}$) and Bob
(owning mode $m_{2}$) can implement the same Gaussian measurements on their
shared state $\hat{\varrho}_{m_{1}m_{2}}$, however, obtain contradictory
outcomes \cite{Handchen}. Essentially, Alice can convince Bob (who does not trust her) that
their shared state $\hat{\varrho}_{m_{1}m_{2}}$ is entangled, while the
converse is impossible. The most interesting application of asymmetric
steering is that it provides security in 1SDI-QKD, where the measurement
apparatus of one party only is untrusted.

\begin{figure}[th]
\centerline{\includegraphics[width=0.4\columnwidth,height=4cm]{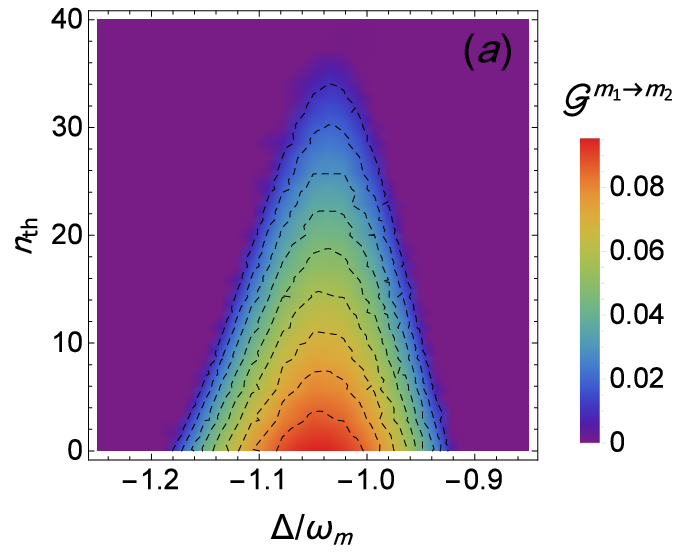}
		\includegraphics[width=0.4\columnwidth,height=4cm]{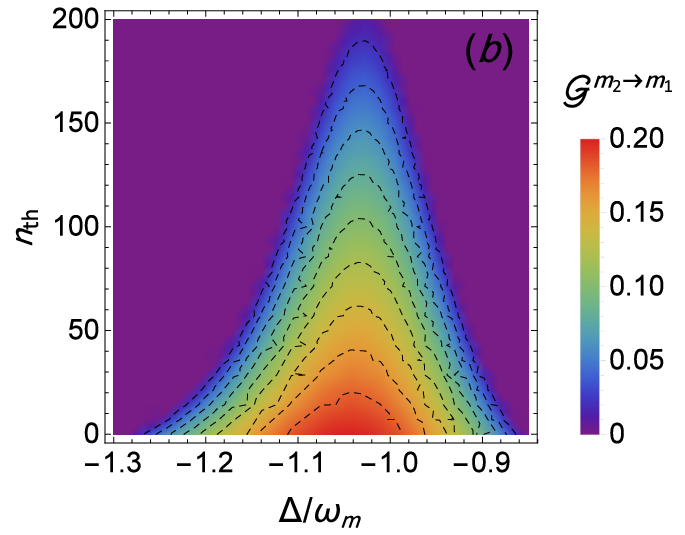}}
\centerline{\includegraphics[width=0.4\columnwidth,height=4cm]{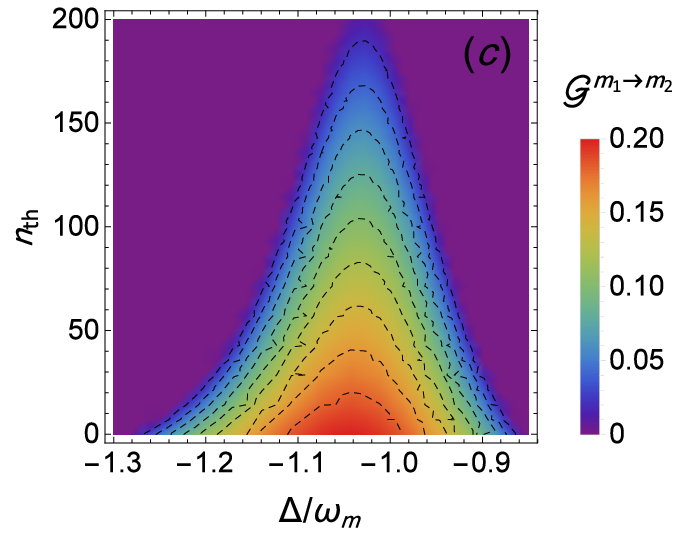}
		\includegraphics[width=0.4\columnwidth,height=4cm]{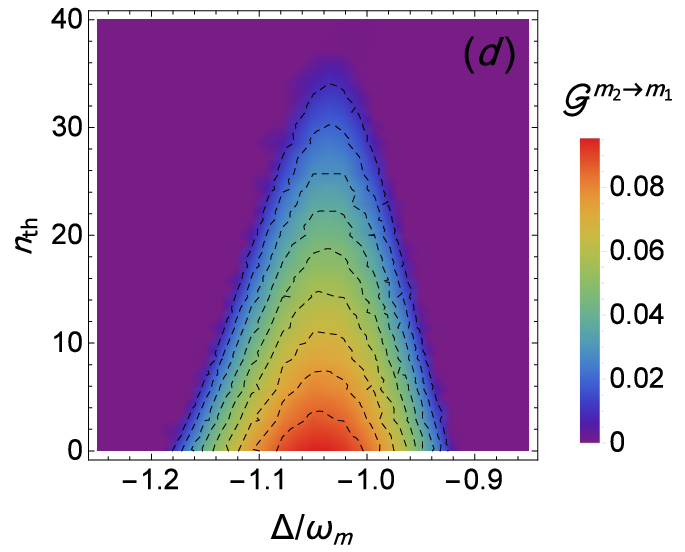}}
\caption{(a) and (b) show, respectively, $\mathcal{G}^{m_{1}\rightarrow
m_{2}}$ and $\mathcal{G}^{m_{2}\rightarrow m_{1}}$ versus the normalized
detuning $\Delta /\protect\omega _{m}$ and the common mean thermal phonon
number $n_{\text{th},1}=n_{\text{th},2}=n_{\text{th}}$ for $G_{1}/\protect%
\omega_{m}=0.5$ and $G_{2}/\protect\omega_{m}=0.4$. (c) and (d) show,
respectively, $\mathcal{G}^{m_{1}\rightarrow m_{2}}$ and $\mathcal{G}%
^{m_{2}\rightarrow m_{1}}$ versus $\Delta /\protect\omega _{m}$ and $n_{%
\text{th}}$ for $G_{1}/\protect\omega_{m}=0.4$ and $G_{2}/\protect\omega%
_{m}=0.5$. The other parameters are $\protect\omega _{m}/2\protect\pi =947~%
\mathrm{kHz}$, $\protect\kappa /2\protect\pi =215~\mathrm{kHz}$, $\protect%
\gamma /2\protect\pi =140~\mathrm{Hz}$, and $\protect\zeta /\protect\omega %
_{m}=0.1$.}
\label{f4}
\end{figure}

Next, we continue by analysing the steerabilities $\mathcal{G}%
^{m_{1}\rightarrow m_{2}}$ and $\mathcal{G}^{m_{2}\rightarrow m_{1}}$ under
influence of the common mean thermal phonon number $n_{\mathrm{th,1}}=n_{%
\mathrm{th,2}}=n_{\mathrm{th}}$ and the normalized detuning $\Delta /\omega
_{m} $ for $\zeta /\omega _{m}=0.1$. In [Figs. \ref{f4}(a) and \ref{f4}(b)],
we used $G_{1}/\omega_{m}=0.5$ and $G_{2}/\omega_{m}=0.4$. While in [Figs. %
\ref{f4}(c) and \ref{f4}(d)], we used $G_{1}/\omega_{m}=0.4$ and $%
G_{2}/\omega_{m}=0.5$. As can be seen, $\mathcal{G}^{m_{1}\rightarrow m_{2}}$
and $\mathcal{G}^{m_{2}\rightarrow m_{1}}$ are maximum close to $n_{\mathrm{%
th}}=0$ and $\Delta /\omega _{m}\approx -1$. However, they decrease by
increasing $n_{\mathrm{th}}$, exhibiting rather different trends against
thermal noise.

\begin{figure}[t]
\centerline{\includegraphics[width=0.4\columnwidth,height=4cm]{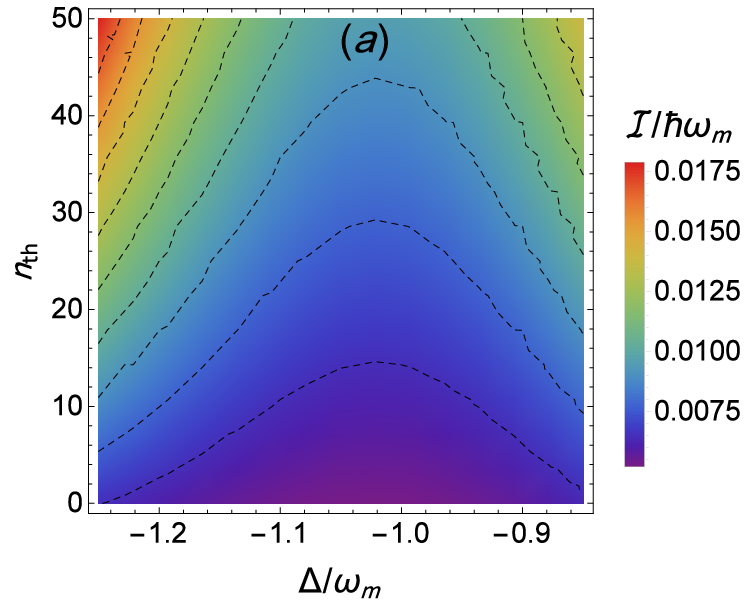}%
\includegraphics[width=0.4\columnwidth,height=4cm]{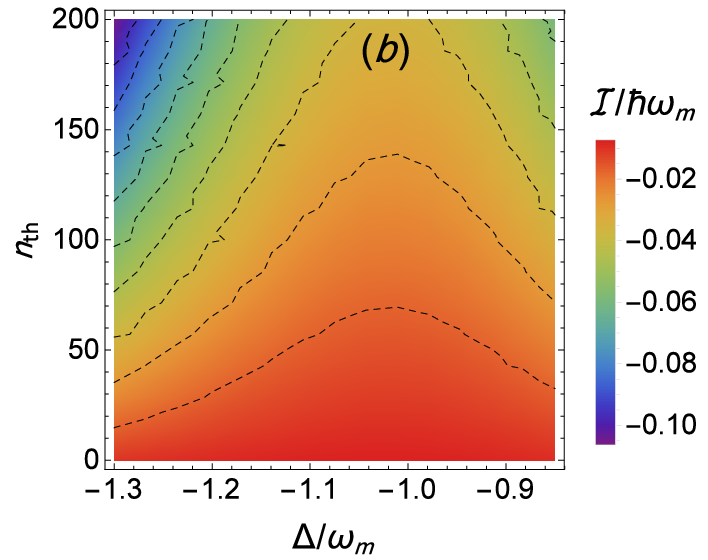}}
\caption{Plot of $\mathcal{I}$, in units of $\hbar\protect\omega_{m}$,
versus $\Delta/\protect\omega_{m}$ and $n_{\text{th}}$. The parameters in
(a) and (b) are, respectively, the same as in [Figs. \protect\ref{f4}(a) and
\protect\ref{f4}(b)] and [Figs. \protect\ref{f4}(c) and \protect\ref{f4}%
(d)]. }
\label{f5}
\end{figure}

Strikingly, Fig. \ref{f4} shows that thermal noise, not only reduces the
steerabilities $\mathcal{G}^{m_{1}\rightarrow m_{2}}$ and $\mathcal{G}%
^{m_{2}\rightarrow m_{1}}$, but may play a positive role in inducing and
controlling one-way steering by the mediation of the ratio $G_{1}/G_{2}$.
Indeed, in [Figs. \ref{f4}(a) and \ref{f4}(b)], where $G_{1}/G_{2}>1$, the
steering $\mathcal{G}^{m_{1}\rightarrow m_{2}} $ disappears completely for $%
n_{\mathrm{th}}>35$. In contrast, $\mathcal{G}^{m_{2}\rightarrow m_{1}}$ can
be detected even for $n_{\mathrm{th}}\approx200$, meaning that the state $%
\hat{\varrho}_{m_{1}m_{2}}$ is one-way steerable from $m_{2}\rightarrow
m_{1} $ for $35<n_{\text{th}}<200$. While, in [Figs. \ref{f4}(c) and \ref{f4}%
(d)], where $G_{1}/G_{2}<1$, $\mathcal{G}^{m_{2}\rightarrow m_{1}}$ vanishes
completely for $n_{\mathrm{th}}>35$. In contrast $\mathcal{G}%
^{m_{1}\rightarrow m_{2}}$ still persist and can be detected for $n_{\mathrm{%
th}}\approx200$, which means that the state $\hat{\varrho}_{m_{1}m_{2}}$ is
one-way steerable from $m_{1}\rightarrow m_{2}$ for $35<n_{\text{th}}<200$.
In fact, [Figs. \ref{f4}(a) and \ref{f4}(b)] show that Bob could stop Alice
to steer his state by adding thermal noise to his mode $m_{2}$ provided that
$G_{1}/G_{2}<1$ while he still can steer Alice's mode $m_{1}$. The same
operation can also be accomplished by Alice in the reverse direction as
illustrated in [Figs. \ref{f4}(c) and \ref{f4}(d)].

Finally, Fig. \ref{f5}(a) shows the difference $\mathcal{I}$ under the same
circumstances of [Figs. \ref{f4}(a) and \ref{f4}(b)], while Fig. \ref{f5}(b)
shows $\mathcal{I}$ under the same circumstances of [Figs. \ref{f4}(c) and %
\ref{f4}(d)]. Also, we remark that the direction of one-way steering depends
on the sign of $\mathcal{I}$, i.e., in Fig. \ref{f5}(a) where $\mathcal{I}>0$%
, one-way steering emerges from $m_{2}\rightarrow m_{1}$ in [Figs. \ref{f4}%
(a) and \ref{f4}(b)]. Whereas, in Fig. \ref{f5}(b) where $\mathcal{I}<0$,
one-way steering emerges through the reverse direction $m_{1}\rightarrow
m_{2}$ in [Figs. \ref{f2}(c) and \ref{f2}(d)]. These observations consist
with the results illustrated in Figs. \ref{f2} and \ref{f3}.

In [Figs. \ref{f2}(c) and \ref{f2}(d)] or [Figs. \ref{f4}(c) and \ref{f4}%
(d)], Alice's manipulation can be viewed as part of a legal step in 1SDI-QKD
protocol \cite{qcry}, where she can get one-way steering by performing local
measurements on her mode $m_{1}$(e.g., adding thermal noise). While, in
[Figs. \ref{f2}(a) and \ref{f2}(b)] or [Figs. \ref{f4}(a) and \ref{f4}(b)],
Bob can obtain one-way steering in the reverse direction by performing local
measurements on mode $m_{2}$, where his manipulation can be regarded either
as a legitimate step or an adversarial attack in 1SDI-QKD protocol \cite%
{qcry}. From this perspective, the presented one-way steering manipulation
scheme can be connected with the 1SDI-QKD protocol, knowing that the security of such protocol depends fundamentally on the direction of steering \cite{qcry}.

\section{Experimental detection of the generated optomechanical steering \label{sec4}}

A challenging aspect of any proposal involving quantum correlations creation
between mechanical degrees of freedom is the actual experimental detection
of the generated quantum correlations. Here, we discuss how the generated
mechanical steerabilities $\mathcal{G}^{m_{1}\rightarrow m_{2}}$ and $%
\mathcal{G}^{m_{2}\rightarrow m_{1}}$ can be measured experimentally. For
two optical modes, measuring their nonlocal correlations can be
straightforwardly accomplished by the well-developed method of homodyne
measurement \cite{SHD}. In our case of two mechanical modes, the situation
is less straightforward. However, the measurement can be indirectly
performed by transferring adiabatically the quantum correlations from the
two mechanical modes back to two auxiliary optical modes \cite%
{palomaki,vitali}. Indeed, we assume that the $j $\textrm{th} mechanical
mirror is perfectly reflecting in both sides, then a fixed and partially
transmitted mirror can be placed instead it to form another Fabry-P\'{e}rot
cavity. We denote by $c_{j}$ the $j$\textrm{th} intracavity auxiliary mode
whose annihilation operator, dissipation rate, and effective detuning are $%
\hat{c}_{j}$, $\kappa _{c_{j}}$, and $\Delta_{c_{j}}$, respectively.
Introducing the $j\text{th}$ mechanical annihilation operator $\delta \hat{m}%
_{j}=(\text{i}\delta \hat{p}_{j}+\delta \hat{q}_{j})/\sqrt{2}$, one can show
that the fluctuation operator $\delta \hat{c}_{j}$ obeys an equation
analogous to Eq. (\ref{e7}), i.e.,
\begin{equation}
\frac{d}{dt}\delta \hat{c}_{j}=-\left( \kappa _{c_{j}}+\mathrm{i}\Delta
_{c_{j}}\right) \delta \hat{c}_{j}+\mathrm{i}\frac{G_{c_{j}}}{2}(\delta \hat{%
m}_{j}^{\dag }+\delta \hat{m}_{j})+\sqrt{2\kappa _{c_{j}}}\hat{c}_{j}^{in}%
\text{,}  \label{e22}
\end{equation}%
where $G_{c_{j}}$ is the effective coupling between the modes
$c_{j}$ and $m_{j}$, with $\hat{c}_{j}^{in}$ being the vacuum radiation input noise acting on mode $c_{j}$.

Furthermore, assuming that the auxiliary mode is driven by a weak laser
field, i.e., $\left\vert \langle \hat{o}_{j}\rangle \right\vert \gg
\left\vert \langle \hat{c}_{j}\rangle \right\vert $, then the back-action of
the auxiliary mode $c_{j}$ on the $j$\textrm{th} mirror can be safely
neglected. Hence, the presence of the auxiliary mode cannot alter the system
dynamics governed by Eqs. (\ref{e2}) and (\ref{e3}). Now, if we choose
parameters so that $\Delta _{c_{j}}=\omega _{m}\gg $ $\kappa _{c_{j}}$,
which is relevant for quantum-state transfer \cite{vitali}, Eq. (\ref{e22})
can be simplified by means of the rotating wave approximation \cite{elqars},
so to get%
\begin{equation}
\frac{d}{dt}\delta \tilde{\hat{c}}_{j}=-\kappa _{c_{j}}\delta \tilde{\hat{c}}%
_{j}+\mathrm{i}\frac{G_{c_{j}}}{2}\delta \tilde{\hat{m}}_{j}+\sqrt{2\kappa
_{c_{j}}}\tilde{\hat{c}}_{j}^{in},  \label{e23}
\end{equation}%
where $\tilde{\Xi}=\Xi e^{i\omega _{m}t}$ for $\Xi\equiv \{\delta \hat{c}%
_{j}, \delta \hat{m}_{j}$, $\hat{c}_{j}^{in}\}$. An optimal quantum-state
transfer, from the mechanical mode $m_{j}$ to the auxiliary mode $c_{j}$,
can be achieved when the later follows adiabatically the former, which
requires $\kappa _{c_{j}}\gg G_{c_{j}} $ \cite{cat}. Hence, by setting $%
\frac{d}{dt}=0$ in Eq. (\ref{e23}) and using $\tilde{\hat{c}}_{j}^{out}=%
\sqrt{2\kappa _{c_{j}}}\delta \tilde{\hat{c}}_{j}-\tilde{\hat{c}}_{j}^{in}$
\cite{Milburn}, we finally get
\begin{equation}
\tilde{\hat{c}}_{j}^{out}=\frac{\mathrm{i}G_{c_{j}}}{\sqrt{2\kappa _{c_{j}}}}%
\delta \tilde{\hat{m}}_{j}\ +\tilde{\hat{c}}_{j}^{in}.  \label{e24}
\end{equation}

The above equation shows that the mechanical covariance matrix (\ref{e19})
can be fully reconstructed via homodyning the output auxiliary fields using
a single homodyne detector \cite{SHD}. This therefore allows us the
numerical estimation of the mechanical steerabilities $\mathcal{G}^{m_{1}
\rightarrow m_{2}}$ and $\mathcal{G}^{m_{2} \rightarrow m_{1}}$.
\section{Conclusions \label{sec5}}

In a double-cavity optomechanical system, stationary Gaussian quantum
steering between two spatially separated mechanical modes is studied. The
two cavities are coupled via an optical fiber, and driven by blue detuned
lasers within the unresolved sideband regime. Using realistic experimental
parameters, we showed that strong asymmetric steering can be achieved
between the two considered modes. Furthermore, we showed that the direction
of one-way steering can practically be controlled through the powers of the
drive lasers and the temperatures of the cavities. This therefore provides a
more flexible and feasible way to manipulate the direction of one-way
steering in experimental operations. Besides, we revealed that the direction
of one-way steering depends on the sign of the difference between the mean
energies of the mechanical modes. Also, we discussed how the generated
steering can be measured experimentally. The presented one-way steering
manipulation scheme can be connected with the 1SDI-QKD protocol, knowing that the
security of such protocol depends crucially on the direction of steering.

The feasibility of our proposal is verified using realistic experimental
parameters within the unresolved sideband regime which makes the obtained
results closer to the experimental reality. In addition, our scheme does not
require additional squeezed light to drive the two cavities, which partly
reduces the experimental realization requirement, and partly makes the
presented scheme a very promising candidate for building up a tabletop
network of optomechanical nodes connected by optical fibers. In practice, the losses---caused by absorption, scattering, etc---within the fiber couplers over long distance, can be overcome by using very-long ultra-low-loss fiber tapers \cite{Brambilla} or optical amplifier technique based on devices that boost the signal power of the optical fiber without converting it to electrical signal \cite{OAT}.

This work may be meaningful for the distribution of long-distance trust which is of fundamental importance in secure quantum communication.

\section*{Data Availability Statement}

No Data associated in the manuscript

\end{document}